\documentclass[11pt]{article}
\usepackage{hyperref}
\usepackage{amssymb}
\usepackage{amsfonts}
\usepackage{amsmath,amsthm}
\usepackage{graphics,epsfig}
\usepackage{xcolor}
\usepackage{subfigure}
\usepackage{graphicx,epsfig, color }
\usepackage{cite}
\usepackage{float}
\usepackage{booktabs}
\usepackage{multirow}
\usepackage{perpage}
\usepackage{subfigure}
\usepackage{caption}
\date{}

\newcommand{\dd}{\mathrm{d}}

\begin{document}
	
	\setcounter{page}{1}
	\pagestyle{plain}

	\begin{center}
		\Large{\bf  Brans-Dicke scalar-tensor gravity and unimodular FRW
			cosmology with Coleman-Weinberg
			potential}\\
		\small \vspace{1cm} {\bf Manda Malekpur\footnote{ E-mail address: manda\_malekpour@semnan.ac.ir}\quad and \quad{\bf Hossein Ghaffarnejad\footnote{ hghafarnejad@semnan.ac.ir (Corresponding Author)}}}
		\\
		\vspace{0.25cm}
		Faculty of Physics, Semnan University, P.C.35131-19111, Semnan, Iran
	\end{center}
	
	\begin{abstract}
		As an alternative gravity theories with respect to General Relativity, the Brans-Dicke scalar-tensor model is well known, in which the Brans-Dicke parameter the controls the best fit observational, and so its cosmological model is studied in the literature well. On the other side, the Coleman-Weinberg self-interaction scalar field potential coming from radiation corrections of Feynman diagrams is used to describe the best fit cosmic inflation phase and reheating phase. We study inflation in unimodular Brans–Dicke gravity with a Coleman–Weinberg potential. We investigate the observational constraints on the Brans–Dicke parameter \(\omega\) and identify its allowed range in light of current cosmological data. Also, in the slow--roll regime, we derive inflation parameters and study the scalar spectral index $n_s$ and the tensor-to-scalar ratio $r$ in terms of the model parameters, and determine the region of the parameter space allowed by observations. Comparing these predictions with the P-ACT-LB-BK18 constraints, we find that the model is consistent with the data at the $95\%$ CL level for $77\lesssim\omega\lesssim1148$ and within the $ 68\% $ CL for $134\lesssim\omega\lesssim272$.
	\end{abstract}
	
	\section{Introduction}\label{sec1}
	The classical cosmology faced several problems, such as the flatness, horizon, and baryon asymmetry problems, among others, which led to the proposal of inflationary models to overcome these shortcomings	\cite{starobinsky1979relict,fakir1990improvement,liddle1994formalizing,Rio02,wang2014inflation,senatore2016lectures,baumann2018tasi}.
	To
	address the puzzles of the Standard Big Bang Cosmology,
	\textquotedblleft inflationary cosmology\textquotedblright\ proposed
	by Alan Guth in 1981 introduced a period of nearly exponential
	expansion in the early universe \cite{guth1981inflationary}. In this
	stage, quantum fluctuations are generated to provide seeds for
	fluctuations and anisotropy of the Cosmic Microwave Background (CMB)
	\cite{Garcia-Bellido:2011kqb}. In this scenario, a scalar field
	known as the \textquotedblleft inflaton\textquotedblright\ slowly goes down the minimum of its
	potential under the slow--roll condition. Present observational data
	show that this paradigm is successful \cite{A.D.Linde}. 
	In recent decades, modified gravity theories have attracted more
	attention in cosmology. In particular, Unimodular Gravity (UG) is
	one of the simplest modifications of General Relativity (GR). One of
	the main unresolved issues in modern physics is the \textquotedblleft cosmological
	constant\textquotedblright\ problem, which arises from the huge difference (about 60
	to 120 orders of magnitude) between theoretical predictions of
	vacuum energy density and its observed value \cite{Nojiri:2017ncd}.
	In the first attempt, Einstein formulated the trace-free
	gravitational field equations in 1919 \cite{einstein1952principle},
	which later became associated with UG. By
	the way, his work does not relate to the cosmological constant, and
	his research focuses on the structure of point particles in GR
	\cite{jain2012testing}. Several decades later, in 1970, Anderson and
	Finkelstein explored this idea \cite{anderson1971cosmological} and
	they were getting a lot of attention in last years
	\cite{gao2014cosmological,cho2015unimodular,barvinsky2017darkness,barvinsky2019inflation,barvinsky2019dynamics,leon2022inflation, Malekpour:2023lsf, Nozari:2024hip, Malekpour:2025baf}. This theory
	imposes a constraint on the determinant of the metric $\sqrt{-g}=1
	$, under which the nature of the cosmological constant shifts from a
	fundamental parameter to an integration constant of the field
	equations
	\cite{weinberg1989cosmological,unruh1989unimodular,liddle1993end}.
	Therefore, the field equations in UG framework are different from
	those in general relativity. Although UG and GR in level classically
	are equivalent, in the quantum regime and at high energies, they can
	be compared with observational data \cite{Plaza:2025nip}. One of the
	interesting features of UG is that the cosmological constant appears
	naturally as an integration constant. While in the standard Einstein
	field equations it was added manually into the field equations. In
	light of this, it has been proposed that UG may provide a way to
	resolve the large difference between the theoretically predicted and
	the observed values of the cosmological constant in GR. Another
	feature of UG is that it reformulates the
	fine--tuning problem of the cosmological constant
	\cite{jain2012cosmological,Nojiri:2016ygo,Odintsov:2016imq}. Despite
	the extensive discussions and success of UG, as Weinberg mentions in
	his article, the cosmological constant problem is still unresolved
	\cite{weinberg1989cosmological}.
	
	As we know, GR is a successful theory in the weak gravity regime
	\cite{Ozer:2021qjb}. It has numerous successes, such as black holes,
	gravitational redshift, perihelion precession
	of Mercury, cosmology, and gravitational waves. However, it was clear that GR does not include Mach's principle. The theory of GR is a geometric theory whose fundamental dynamical variable is the metric tensor variable describing the structure of spacetime. In contrast, in scalar-tensor theories, gravity is
	described by the metric tensor and a dynamical scalar field. In this
	approach, to incorporate Mach's principle into GR, several authors
	investigated and studied \cite{Faraoni:1999hp, Fujii:2003pa, Brans:2005ra}. The first attention to incorporate this principle led to the proposal of scalar-tensor theories, the most important and simplest of which was the Brans--Dicke (BD) theory in 1961 \cite{Brans:1962zz}, in which a scalar field is non-minimally coupled to Ricci curvature \cite{Tirandari:2017nzy}.
	
	After the universe's  accelerated expansion was found in 1998, scalar-tensor theory, and especially BD theory, was suggested to explain the cosmic
	expansion phenomenon. Nevertheless, several analyses have been conducted on inflationary models in the BD theory \cite{Mau, Roy:2017mnz, Sharma:2019okc, Almeida:2022qld}.
	In BD theory, the parameter $ \omega $ controls deviations from GR. Its observational bounds, however, depend on the scale probed. At local scales, the Cassini mission yields a stringent constraint, $\omega \gtrsim 10^{4}$ \cite{Bertotti:2003rm}, whereas cosmological data from CMB and large-scale structure gives a much weaker bound $\omega \gtrsim10^{3}$ \cite{Avilez:2013dxa, Joudaki:2020shz}. This difference is naturally explained in theories with screening mechanisms, where the scalar field is suppressed in dense environments but remains active on cosmological scales \cite{Sletmoen:2024are}. 
	As noted in \cite{Tahmasebzadeh:2016irh}, if the BD scalar field decays after inflation, the subsequent cosmology reduces to GR and the present-day Solar System bound does not apply to the inflationary era.
	As a result, the local bound does not necessarily apply to the inflationary epoch, which occurs at extremely low densities. In this work, we therefore treat $ \omega $ as a free parameter during inflation and examine whether the UBD model can yield observationally viable inflation for reasonable values of $ \omega $. 
	
	In this paper, we present an inflationary model in the original
	formulation of BD theory within the framework of UG. We examine
	whether UBD gravity theory results in an inflationary phase in the
	early universe or not. We use the Coleman--Weinberg (CW) potential in this framework to
	analyze behaviors in the inflationary model. Then, we study the
	observable inflation and derive slow--roll parameter approximations
	for this model. We compare and check results viability in light of
	the ACT datasets at $68\%$ and $95\%$ confidence levels (CL) \cite{ACT:2025tim, ACT:2025fju}. The paper is organized as follows:\\ 
	In Sec. \ref{sec2}, we introduce the action of BD gravity in the context of the unimodular framework in the presence of the CW potential, and derive the corresponding field equations for a spatially flat Friedmann–Robertson–Walker (FRW). In Sec. \ref{sec3}, we investigate slow--roll cosmic inflation in the presence of the CW potential, adopting the ansatz \(\Lambda(\phi)=\Lambda_0+A\phi^n\). We then derive the inflationary observables, including the scalar spectral index \(n_s\) and the tensor-to-scalar ratio \(r\). Next in Sec. \ref{sec4}, we present the numerical results for the model and obtain the potential the coupling $ \lambda $. Finally, in Sec. \ref{sec5} we summarize our main findings and discuss. We work in units where $ c = \hbar=1$, $\kappa^{2}\equiv 8 \pi G $ and we use the metric signature $(-, +, +, +)$.\\

%#######################################
		
	\section{Generalized Unimodular Brans-Dicke Gravity}\label{sec2}
	In this section, we consider the Brans--Dicke action in the unimodular framework and derive the corresponding background field equations as fallows
	\begin{equation}\label{eq1}
		S=\int d^{4}x \left\lbrace \sqrt{-g} \left(\phi R- \omega \frac{\nabla_{\nu}\phi \nabla^{\nu}\phi}{\phi}- V(\phi) \right)
		- \Lambda(x^{\mu}) (\sqrt{-g}-1)\right\rbrace \,,
	\end{equation}
	where $\Lambda (x^\mu) $ is the Lagrange multiplier, $
	\omega $ is the BD dimensionless parameter which in limit $\omega\to\infty$
	reaches to the GR \cite{Weinberg}. Next, we consider a well-known CW inflationary potential for the scalar field $\phi$, which is given by
	\begin{equation}\label{eq2}
		V_{cw}(\phi) = \lambda  \left[ \phi^4 \ln\bigg( \frac{\phi}{\phi_0}\bigg) + \frac{\phi^4_0}{4} - \frac{\phi^4}{4}
		\right]\,,
	\end{equation} 
	in which $\lambda>0$ is the coupling constant with
	dimensions of $(mass)^{-4}$ and $\phi_0$ is a particular value of the
	scalar field for which the potential reaches its minimum value as
	$V_{cw}(\phi_0)=V^{\prime}_{cw}(\phi_0)=0$. The above action with
	$V_{cw}=\Lambda=0$, called the BD scalar tensor gravity
	reduces to the Einstein Hilbert action for a constant field $\phi_{G}=1/2\kappa^2=1/16\pi G$.
	This means that $\phi$ is a fluctuation about Newton's gravity
	coupling parameter according to Mach's principal \cite{BD} (see also \cite{CBD}). This kind of potential is suitable for describing cosmic inflation, with the reheating phase beginning when the inflaton field reaches the minimum of the potential at the end of inflation. Furthermore, the CW potential at the minimum point vanishes, and so we can infer that the above gravity model reaches the Einstein-Hilbert action functional at the minimum point $\phi_0=\phi_G$. The Friedmann equations can be derived by varying the action \eqref{eq1} with respect to $ g_{\mu\nu} $ as follows
	    \begin{align}\label{eq3}
		R_{\mu\nu} - \frac{1}{2} g_{\mu\nu} R& = \frac{\omega}{\phi^{2}} (\nabla_{\mu}\phi \nabla_{\nu}\phi
		- \frac{1}{2} g_{\mu\nu} \nabla_{\rho}\phi \nabla^{\rho} \phi)\notag\\& - \frac{1}{\phi} ( g_{\mu\nu} \Box \phi
		- \nabla_{\mu} \nabla_{\nu}\phi ) - \frac{V_{cw}(\phi)}{2\phi} g_{\mu\nu}
		- \frac{\Lambda(x^\mu)}{2\phi} g_{\mu\nu}\,.
	\end{align}
	Taking the variation of this action with respect to $ \phi $ gives us the
	equation of motion for the BD scalar field as
	\begin{equation}\label{eq4}
		\Box\phi = \frac{1}{2} \frac{\nabla_{\nu}\phi \nabla^{\nu}\phi}{\phi} - \frac{\phi R}{2 \omega}+ \frac{\phi}{2\omega}
		V_{cw}^{\prime}(\phi),
	\end{equation}
	where we substitute the UG condition $\sqrt{-g}=1$. The d'Alembert
	operator is $\Box \equiv  g^{\mu\nu}\nabla_{\mu}\nabla_{\nu}$ and \textquotedblleft$\prime$\textquotedblright\ means derivative with respect to the BD field $\phi$. 
Then, we get the trace of Eq. \eqref{eq3}, which gives
\begin{equation}\label{eq5}
		R = \frac{1}{\phi}\Big(3\Box\phi +\frac{\omega}{\phi}\nabla_{\kappa}\phi \nabla^{\kappa}\phi+ 2V_{cw}(\phi) + 2\Lambda(x^\mu) \Big)\,,
	\end{equation}
	then applying the above equation in \eqref{eq4}, we obtain
	\begin{equation}\label{eq6}
		(2\omega+3)\Box\phi-\phi V_{cw}^\prime (\phi)+2V_{cw}(\phi)+2\Lambda(x^\mu)=0\,.
	\end{equation} 
	
We consider the universe to be described by the spatially flat, homogeneous and isotropic FRW metric with a time-dependent scale factor $ a(t) $ as follows
\begin{equation}\label{eq7}
	ds^{2} = -dt^{2}+ a^{2}(t) \sum_{i=1}^{3} \left( dx^{i}\right)^{2}\,,
\end{equation}
that this metric does not satisfy the unimodular constraint $\sqrt{-g}=1$. Therefore, we introduce a new time coordinate defined by $ d\tau=a^{3}(t)\ dt$. Consequently, the spatially flat unimodular FRW metric \eqref{eq7} can be rewritten as follows
	\begin{equation}\label{eq8}
		ds^{2} = -a^{-6} \left( \tau\right)  d\tau^{2} + a^{2}\left( \tau\right)  \sum_{i=1}^{3} \left( dx^{i}\right)^{2}\,,
			\end{equation}
where $ \tau $ is unimodular cosmic time coordinate, and it is obvious
that the determinant of the metric field is unity. Now, using the above metric, we calculate the non-vanishing components of the Ricci tensor and the Ricci scalar, which are given by
\begin{equation}\label{eq9}
		R_{\tau\tau}= -3\dot{\mathcal{H}} - 12 \mathcal{H}^2\,,
	\quad R_{ij}= a^8\left( \dot{\mathcal{H}} + 6\mathcal{H}^2 \right)  \delta_{ij}\,,
	\quad R= a^6 \left( 6 \dot{\mathcal{H}} +30 \mathcal{H}^2\right) \,,
\end{equation}
where $\mathcal{H}= \frac{1}{a}\frac{da}{d\tau}$ is the alternative
Hubble parameter in the UG framework. The relation between $\mathcal{H}$ and the standard Hubble parameter defined with respect to the proper cosmic time is $H=a^3\mathcal{H}$ and dot $ ( \, \dot{} = \frac{d}{d\tau})$ stands for a derivative with respect to the time parameter $\tau$. 
Then by applying Eqs. \eqref{eq8} and \eqref{eq9}, the $\tau \tau$ and $ ij $ components of the Friedmann equations \eqref{eq3} are respectively
\begin{equation}\label{eq10}
3\mathcal{H}^2=\frac{\omega}{2}\bigg(\frac{\dot{\phi}}{\phi}\bigg)^2-3\mathcal{H}\frac{\dot{\phi}}{\phi}+\frac{a^{-6}(V_{cw}+\Lambda)}{2\phi}\,,
\end{equation} 
and
\begin{equation}\label{eq11}
	2\dot{\mathcal{H}}+9\mathcal{H}^2=-\frac{\omega}{2}\bigg(\frac{\dot{\phi}}{\phi}
	\bigg)^2-\frac{\ddot{\phi}}{\phi} -5\mathcal{H}\frac{\dot{\phi}}{\phi}+\frac{a^{-6}(V_{cw}+\Lambda)}{2\phi}\,.
\end{equation}	
	 For simplicity, we omit the explicit dependence of \(\Lambda\) on \(x^\mu\) and \(V_{\mathrm{cw}}\) on \(\phi\). Substituting the line element \eqref{eq7}, the Klein-Gordon equation \eqref{eq4} can be expressed by
	\begin{equation}\label{eq12}
		\frac{\ddot \phi}{\phi}+6 \mathcal{H}
		\frac{\dot{\phi}}{\phi}-\frac{1}{2}\frac{\dot\phi^{2}}{\phi^2}-\frac{1}{\omega}\bigg[3(\dot{\mathcal{H}}+5\mathcal{H}^2)-\frac{
			V'_{cw}} {2\ a^{6}}\,\bigg]=0\,,
	\end{equation}
	Combining Eq.s \eqref{eq10} and \eqref{eq11}, so we have
	\begin{equation}\label{eq13}
	2 \dot{\mathcal{H}} + 6 \mathcal{H}^2 + \omega(\frac{\dot{\phi}}{\phi})^{2}	+2 \mathcal{H}\, (\frac{\dot{\phi}}{\phi}) + \frac{\ddot{\phi}}{\phi}=0\,.
    \end{equation}
This equation plays an important role in the unimodular reconstruction framework. Let us consider a universe described by a power-law scale factor of the following form \cite{Nojiri:2015sfd}
\begin{equation}\label{eq14}
	a\left(t\right)= \left( \frac{t}{t_{0}}\right)^{b}   \Longrightarrow H = \frac{b}{t} \,,
\end{equation}
where ${t_{0}} $ and $ b $ are constants, with $ b $ defined as $ b = 2 / [ 3(1 + \omega_0 ) ] $, where $ \omega_0 $ denotes the equation-of-state parameter. By introducing the new time coordinate $ (d\tau=a^3(t)\ dt) $, we can easily obtain
\begin{equation}\label{eq15}
	\tau = \frac{t_{0}}{3b+1} \left( \frac{t}{t_{0}}\right)^{3b+1}\,,
\end{equation}
	and inserting Eq. \eqref{eq15} in Eq. \eqref{eq14}, we find the scale factor in terms of the variable $\tau$ as follows
	\begin{equation}\label{eq16}
		a\left( \tau\right) = \left( \frac{\left(3b+1\right)  \tau}{t_{0}} \right)^{\frac{b}{3b+1}}\,.
	\end{equation}
	Now considering the scale factor as 
	\begin{equation}\label{eq17}
		a\left(\tau\right)= \left( \frac{\tau}{\tau_{0}}\right)^{q}   \Longrightarrow \mathcal{H}= \frac{q}{\tau}\,,
	\end{equation}
	where $q$ and $\tau_{0}$ are arbitrary constants. Hence, the unimodular FRW metric \eqref{eq8} takes form 
	\begin{equation}\label{eq18}
		ds^{2} = -\left( \frac{\tau}{\tau_{0}} \right)^{-6q} d\tau^{2} + \left( \frac{\tau}{\tau_{0}} \right)^{2 q} \sum_{i=1}^3 \left( dx^{i}\right)^{2}\,.
	\end{equation}
A de Sitter expansion is obtained for $q=\frac{1}{3}$. For $\frac{1}{4} \leq q<\frac{1}{3}$, the cosmic evolution corresponds to a quintessence-dominated phase. The cases for $q=\frac{2}{9}$ and $q=\frac{1}{5}$ correspond to dust and radiation dominated universes, respectively. Next step, Eq. \eqref{eq13} can be written as
\begin{equation}\label{eq19}
\frac{\ddot{\phi}}{\phi}
+\omega\left(\frac{\dot{\phi}}{\phi}\right)^2
+\frac{2q}{\tau}\left(\frac{\dot{\phi}}{\phi}\right)
+\frac{2q(3q-1)}{\tau^2}=0\,,
\end{equation} 
and general solution of Eq. \eqref{eq19} is equal to
\begin{equation}\label{eq20}
	\phi(\tau)=\Big[C_1\Big(\tau\Big)^{s_+}+C_2\Big(\tau\Big)^{s_-}\Big]^{\frac{1}{1+\omega}},
	\quad
	s_\pm=\tfrac12\Big[(1-2q)\pm\sqrt{(2q-1)^2-8q(3q-1)(1+\omega)}\Big]\,.
\end{equation}
where $ s_\pm $ are constants of integration. Using Eq. \eqref{eq10} and the solution  \eqref{eq20}, we derive the unimodular Lagrange multiplier as fallows
\begin{equation}\label{eq21}
		\Lambda(\tau)=2\,\tau^{6q-2}\,S(\tau)^{\frac{1}{1+\omega}}
		\left[\,3q^{2}+3q\,L(\tau)-\frac{\omega}{2}\,L(\tau)^{2}\right]-V_{cw}\!\left(S(\tau)^{\frac{1}{1+\omega}}\right)\,,
	\end{equation}
where
\begin{equation}\label{eq22}
	S(\tau)\equiv C_1\,\tau^{s_+}+C_2\,\tau^{s_-}\,,\qquad
	s_\pm=\frac12\left[(1-2q)\pm\sqrt{(2q-1)^2-8q(3q-1)(1+\omega)}\right]\,, 
\end{equation}
	\begin{equation}
		 L(\tau)\equiv\frac{\tau\dot\phi}{\phi}=\frac{s_+C_1\tau^{s_+}+s_-C_2\tau^{s_-}}{(1+\omega)\,S(\tau)}\,.
	\end{equation}
This result is that the cosmological parameter is a time--varying quantity in the UG framework, which supports comic inflation. Now we investigate slow--roll conditions have the best fit with the observational data for the above power-law inflation solution. Furthermore, the continuity equation define as
\begin{equation}\label{eq.rho}
	\dot\rho_{\rm eff} + 3 (\rho_{\rm eff} +P_{\rm eff})=0\,, 
\end{equation}
where $ \rho_{\rm eff} (\phi)$ and $ P_{\rm eff} (\phi) $ are the effective energy density and pressure. Using the relation $ G^\tau_{\tau}=-3a^6\mathcal{H}^2=-\rho_{\rm eff} $ and $G^i_j=-a^6(2\mathcal{\dot{H}}+9\mathcal{H}^2)=P_{\rm eff}\delta^i_j $, and by using the field Eqs. \eqref{eq10} and \eqref{eq11} that behaves as a perfect fluid whose effective density and pressure are give respectively by
\begin{align}\label{rho}
	&\rho_{\rm eff}(\phi)=-T^\tau_\tau=\bigg(\frac{V_{cw}+\Lambda}{2\phi}\bigg)+a^6\bigg(\frac{\omega}{2}\frac{\dot{\phi}^2}{\phi^2}
	-3\mathcal{H}\frac{\dot{\phi}}{\phi}\bigg)\,,\notag\\&
	P_{\rm eff}(\phi)=T^1_1=T^2_2=T^3_3=-\bigg(\frac{V_{cw}+\Lambda}{2\phi}\bigg)+a^6\bigg[\frac{\omega}{2}\frac{\dot{\phi}^2}{\phi^2}+5\mathcal{H}
	\frac{\dot{\phi}}{\phi}+\frac{\ddot{\phi}}{\phi}\bigg]\,.
\end{align}
We use Eq.s \eqref{eq.rho} and \eqref{rho} as follows
\begin{equation}\label{111}
	\frac{(\Lambda^\prime + V^\prime) \dot{\phi}}{2\ \phi} +a^{6} (\frac{ \dot{\phi}}{\phi}) \left[ \omega (\frac{ \ddot{\phi}}{\phi}) + 6\ \omega \mathcal{H} (\frac{ \dot{\phi}}{\phi}) - (\frac{\omega}{2} ) (\frac{ \dot{\phi}}{\phi})^{2} -3 \dot{\mathcal{H}} -15 \mathcal{H}^{2}  \right]=0 \,,
	\end{equation}
where $ \Lambda^\prime = \frac{d\Lambda}{d\phi} $.
In continuation, we use this equations for our purpose in the next section.\\

%###############################################################

\section{Slow--Roll Parameters and Inflationary Observables}\label{sec3}
	We consider Linde setup for the Hubble Slow--Roll approximation in the cosmic inflation \cite{Linde}, \cite{Fakir} such that
	\begin{align}\label{eqslr}
		\left|  \frac{\ddot{\phi}}{\dot{\phi}} \right|  \ll H\,,
	\qquad \left| \frac{\dot{\phi}}{\phi} \right|  \ll H\,,
	\qquad \left| \dot{H}\right| \ll H^{2}\,, \qquad \dot{\phi}^{2} \ll V_{cw}\,.
	\end{align}
			In the UG framework the above slow--roll conditions are transformed to the following forms by using the time transformation between the UG observer $\tau$ and the free falling comoving observer $t$, i.e., $d\tau=a^3dt$ such that
		\begin{align}\label{eq25}
			\frac{d^2\phi}{d\tau^2}\ll-2\mathcal{H}\frac{d\phi}{d\tau},~~~\frac{d\phi}{d\tau}\ll\mathcal{H}\phi,~~~\frac{d\mathcal{H}}{d\tau}\ll-2\mathcal{H}^2,~~
			~\bigg(\frac{d\phi}{d\tau}\bigg)^2 \ll V_{cw}a^{-6}\,.
		\end{align}
		Therefore the slow--roll parameters define as \cite{Linde, Nojir}
	 \begin{equation}\label{eq26}
	 \epsilon_{1} = - \frac{\dot H}{H^{2}}\,, \quad
	 \epsilon_{2} =  \frac{\ddot{\phi}}{H \dot{\phi} }\,, \quad
	 \epsilon_{3} = \frac{\dot{f}\left(\phi\right) }{2 H f\left(\phi\right) }\,, \quad
	 \epsilon_{4} =\frac{\dot{E}}{2 H E}\,,\quad
	 	E\equiv f(\phi)\ \omega+\frac{3 \dot{f}^{2}\left(\phi\right)}{2\kappa^2\dot{\phi}^2}\,,
 	\end{equation}
 	in which the slow--roll parameters satisfy the slow--roll condition $\epsilon_i\ll1 \,; i=1\,, 2\,, 3\,,4$. 
	Moreover, the scalar spectral index \textquotedblleft$ n_s $\textquotedblright \ and the tensor-to-scalar
	ratio \textquotedblleft $r $\textquotedblright \ in terms of the slow--roll parameters are represented respectively as \cite{Nojir}
	\begin{equation}\label{eq27}
		n_s=1-4\epsilon_1-2\epsilon_2+2\epsilon_3-2\epsilon_4\,, \qquad r=\frac{8\,\kappa^2 \, Q_s}{f(\phi)}\,,
		\end{equation} 
	where the parameter $Q_{s}$ is given by
	\begin{equation}\label{eq28}
		Q_{s}= \frac{E \dot{\phi^{2}} }{f\left( \phi\right)  H^{2} (1 +\epsilon_{3} )^{2}}\,.
			\end{equation}
	In the unimodular gravity framework, Eq. \eqref{eq26} takes the following form
	\begin{align}\label{eq29}
		\epsilon_1=-3-\frac{\mathcal{\dot{H}}}{\mathcal{H}^2}\,,\quad
		\epsilon_2=3+\frac{\ddot{\phi}}{\mathcal{H}\dot{\phi}}\,,\quad
		\epsilon_3=\frac{\dot{f}(\phi)}{2\mathcal{H}f(\phi)}\,,\quad
		\epsilon_4=\frac{\dot E}{2\mathcal{H}E}\,,\quad
		E\equiv f(\phi)\ \omega+\frac{3 \dot{f}^{2}\left(\phi\right)}{2\kappa^2\dot{\phi}^2}\,.
	\end{align}
	 Within the framework of UBD gravity and considering in our model, we can rewrite $ E $ as
	 \begin{equation}
	 	E=2\kappa^2(2\omega+3)
	 	\ \Longrightarrow\ \epsilon_4=\frac{\dot E}{2\mathcal{H}E}=0 \,,
	 	\end{equation}
	 and the function $ Q_{s} $ is equal to
	\begin{equation}
		Q_s=\frac{(2\omega+3)\,\dot\phi^2}{\phi \mathcal{H}^2(1+\epsilon_3)^2}\,.
	\end{equation}
Using the above equation with \(\dot{\phi}=2\epsilon_3 \mathcal{H}\phi\), we obtain the tensor-to-scalar ratio as
	\begin{equation}
  r=\frac{16\,(2\omega+3)\,\epsilon_3^2}{(1+\epsilon_3)^2}\,,
	\end{equation}
and the spectral index \eqref{eq27} with $\epsilon_4=0$ is
\begin{equation}\label{eq33}
		n_s=1-4\epsilon_1-2\epsilon_2+2\epsilon_3\,.
		\end{equation}
	With the formulation of our UBD model in place, now we are prepared with a numerical analysis of the model's parameter space to examine its viability and determine the constraints imposed on its parameters by recent observational data.
\\
	\subsection{Inflation with $V_{cw}$ and $\Lambda(\phi)=\Lambda_0+A\,\phi^{n}$}
	
	In this section we consider in which the Lagrange multiplier term is assumed to be to a field-dependent function, $ \Lambda(\phi) $. Therefore, energy–momentum is not strictly conserved. In Eq. \eqref{rho}, we introduce effective energy density and pressure that incorporate the contributions from the BD scalar field \(\phi\) and the unimodular term \(\Lambda\). In our setup, as Eq. \eqref{111} shows, $ \Lambda $ is not a constant.
	Consequently, the effective energy–momentum tensor may satisfy a conservation equation, whereas the energy–momentum associated with the BD scalar field and the variable unimodular parameter \(\Lambda\) is not conserved. This implies that, we take the Lagrange multiplier as the ansatz $\Lambda(\phi)=\Lambda_0+A \,\phi^n$ with $n\in\mathbb{N}$ and $\Lambda_0$, $A$ of arbitrary sign. The contracted Bianchi identity (continuity equation) then fixes the scalar equation as Eq. \eqref{eq12} with $V_{cw}'\to V_{cw}'+\Lambda'$ (which gives the same result as Eq. \eqref{111}), so that $\Lambda(\phi)$ acts as an additional potential term. We start this calculation with auxiliary variables ($x\equiv\phi/\phi_0$):
	\begin{equation}\label{eq34}
	V_{cw}+\Lambda=\lambda\phi_0^4\,W(x)\,,\qquad
	W(x)=x^4\ln x+\frac14-\frac{x^4}{4}+\hat\Lambda+c\,x^n\,,\qquad
	\hat\Lambda=\frac{\Lambda_0}{\lambda\phi_0^4}\,,\quad c=\frac{A\phi_0^{\,n}}{\lambda\phi_0^4}\,,
\end{equation}
and 
\begin{equation}\label{eq35}
	X\equiv\frac{V_{cw}+\Lambda}{a^6\phi \mathcal{H}^2}\,, \qquad
	s\equiv\frac{\phi\,(V_{cw}+\Lambda)'}{V_{cw}+\Lambda}=\frac{xW'(x)}{W(x)}\,. 
\end{equation}
We require a Minkowski vacuum at $x=x_m$, i.e.\ $W(x_m)=W'(x_m)=0$, which fixes
\begin{equation}\label{eq36}
	c=-\frac{4}{n}\,x_m^{\,4-n}\ln x_m,\qquad
	\hat\Lambda=-\hat V_{cw}(x_m)-c\,x_m^{\,n},\qquad \hat V_{cw}(x)=x^4\ln x+\tfrac14-\tfrac14x^4 \,.
\end{equation}
The position of the true minimum is shifted in our model ($ x_m \neq 1 $). Thus, the model is characterized by three free parameters \((n,\omega,x_m)\). In the following, a positive-energy maximum of the effective potential, located before the vacuum (the maximum of the effective potential $W/x^{2}$), is required to achieve a viable slow--roll regime with a small tensor-to-scalar ratio. Our numerical analysis over the parameter $ (x_m) $ shows that this feature is present only for $ (n=1\,, 2\,, 3) $, while no such maximum is found for $ (n\geq4) $. This condition requires \(\hat{\Lambda}\lesssim -1/4\). In the following, we focus on the case \(n=3\), for which \(x_m\) falls within the range \(0.37\lesssim x_m\lesssim 0.47\). Our analysis shows that, for \(n=1\) and \(n=2\), we could not find any parameter values that satisfy the observational constraints. In contrast, for \(n=3\) yields results that are consistent with the observational data, making it a viable choice for the model.
In continue, we can rewrite \eqref{eq29} as fallows
\begin{equation}\label{eq37}
	\frac{\dot\phi}{\phi}=2\epsilon_3 \mathcal{H}\,,\quad
	\dot{\mathcal{H}}=-(3+\epsilon_1) \mathcal{H}^2\,,\quad
	\frac{\ddot\phi}{\phi}=2\epsilon_3(\epsilon_2-3) \mathcal{H}^2\,.
\end{equation}
Substituting the slow--roll parameters \eqref{eq37} in Eqs. \eqref{eq10}-\eqref{eq12}, one
can obtain
\begin{align}
	\tfrac12 X&=3+6\epsilon_3-2\omega\epsilon_3^2  \label{eq38}\,,\\
	\tfrac12 X&=3-2\epsilon_1+2\omega\epsilon_3^2+4\epsilon_3+2\epsilon_3\epsilon_2  \label{eq39}\,,\\
	2\epsilon_3&(3+\epsilon_2-\epsilon_3)+\frac{3(\epsilon_1-2)+2Xs}{\omega}=0  \label{eq40}\,.	
\end{align}
Eliminating $ X $ between \eqref{eq38} and \eqref{eq39}, and then $ \epsilon_1 $ from \eqref{eq40}, we derive the slow--roll parameters and use them to compute as
\begin{align}
	X&=6+12\epsilon_3-4\omega\epsilon_3^2 \label{eq41}\,,\\ 
	\epsilon_1&=\epsilon_3\,(2\omega\epsilon_3-1+\epsilon_2)\label{eq42}\,,\\
	\epsilon_2&=-\frac{(6\omega-3)\epsilon_3+4\omega\epsilon_3^2+\tfrac12Xs-6}{\epsilon_3(2\omega+3)}\,.
\end{align}
During the slow--roll regime ($|\epsilon_{i}|\ll1$ ), we have $X\simeq6$, neglecting the
kinetic terms in \eqref{eq10}, and $\epsilon_2\simeq0$,  which corresponds to $ (\ddot{\phi}\simeq -3H\dot{\phi}) $, we obtain
\begin{equation}\label{eq44}
	\epsilon_3\simeq\frac{2-s}{2\omega+3}
	=\frac{\tfrac12(1-x^4)-2x^4\ln x+2\hat\Lambda+(2-n)\,c\,x^n}{(2\omega+3)\,W(x)}\,,\qquad
	\epsilon_1\simeq\epsilon_3(2\omega\epsilon_3-1)\,,
\end{equation}
In the study of cosmic inflation, another parameter plays a key role in connecting inflationary dynamics with observational constraints. is the number of e-folds. It is defined as 
\begin{equation}
	N=\int_{t_{hc}}^{t_e}H\ \dd t= \int_{\phi_{hc}}^{\phi_e} \frac{H}{\dot{\phi}}\ \dd\phi \,,
\end{equation}
where $t_{hc}$ denotes the horizon-crossing time with
corresponding scalar field value $\phi_{hc}$, at which the CMB pivot scale crosses the Hubble horizon. On the other hand, $t_e$ represents the time at which inflation ends, when the inflaton field reaches the value $\phi_e $. Inflation ends when $\epsilon_1(x_e)=1$. Using the definition given in Eq. \eqref{eq44}, we obtain $\epsilon_3(x_e)=\big(1+\sqrt{1+8\omega}\big)/(4\omega)$. The number of e-folds is then given by
\begin{equation}
	N=\int_{x_{hc}}^{x_e}\frac{\dd x}{2\,x\,\epsilon_3(x)} \simeq\frac{2\omega+3}{2}\int_{x_{hc}}^{x_{e}}\frac{d\ln x}{2-s(x)}\,,
\end{equation}
For $\epsilon_3>0$, the scalar field increases during inflation, leading to $ (x_e>x_{hc}) $. In this case, the field rolls from the hilltop towards the vacuum at $(x_m) $. From the definitions,
$d\ln x/dN=2\epsilon_{3}$ and $d\epsilon_{3}/dN=\epsilon_{3}(\epsilon_{1}+\epsilon_{2}-2\epsilon_{3})$. Hence, we obtain
\begin{equation}\label{eqep3}
	\epsilon_{2}\simeq2\,\frac{d\epsilon_{3}}{d\ln x}-\epsilon_{1}+2\epsilon_{3}
	=2\,\frac{d\epsilon_{3}}{d\ln x}-2\omega\epsilon_{3}^{2}+3\epsilon_{3},\qquad \epsilon_{4}=0 \,.
\end{equation}
Since $\epsilon_{3}=\epsilon_{3}(x)$ along the trajectory, $d\epsilon_{3}/dN=2\epsilon_{3}\,d\epsilon_{3}/d\ln x$; equating this with $\epsilon_{3}(\epsilon_{1}+\epsilon_{2}-2\epsilon_{3})$, and using $\epsilon_{1}\simeq\epsilon_{3}(2\omega\epsilon_{3}-1)$ (the term $\epsilon_{2}\epsilon_{3}$ being second order), gives the expression of $\epsilon_{2}$ above.
By substituting Eq. \eqref{eqep3} in Eq. \eqref{eq33}, the spectral index takes the following form
\begin{equation}
	n_s=1-4\epsilon_1-2\epsilon_2+2\epsilon_3\simeq1-4\omega\epsilon_3^2-4\,\frac{\dd\epsilon_3}{\dd\ln x}\,,
\end{equation}
and tensor-to-scalar ratio is equal
\begin{equation}
	r\simeq16(2\omega+3)\, \epsilon_3^2\simeq\frac{16(2-s)^2}{(2\omega+3)}\,,
\end{equation}
both evaluated at $x_{hc}$ defined by the required $N$. So with these result, now we can determine the allowed ranges in our model.
\\
%####################################################

\section{Observational Constraints on the Brans–Dicke Parameter}\label{sec4}
\begin{figure}[H]
	\centering
	\includegraphics[width=0.8\textwidth]{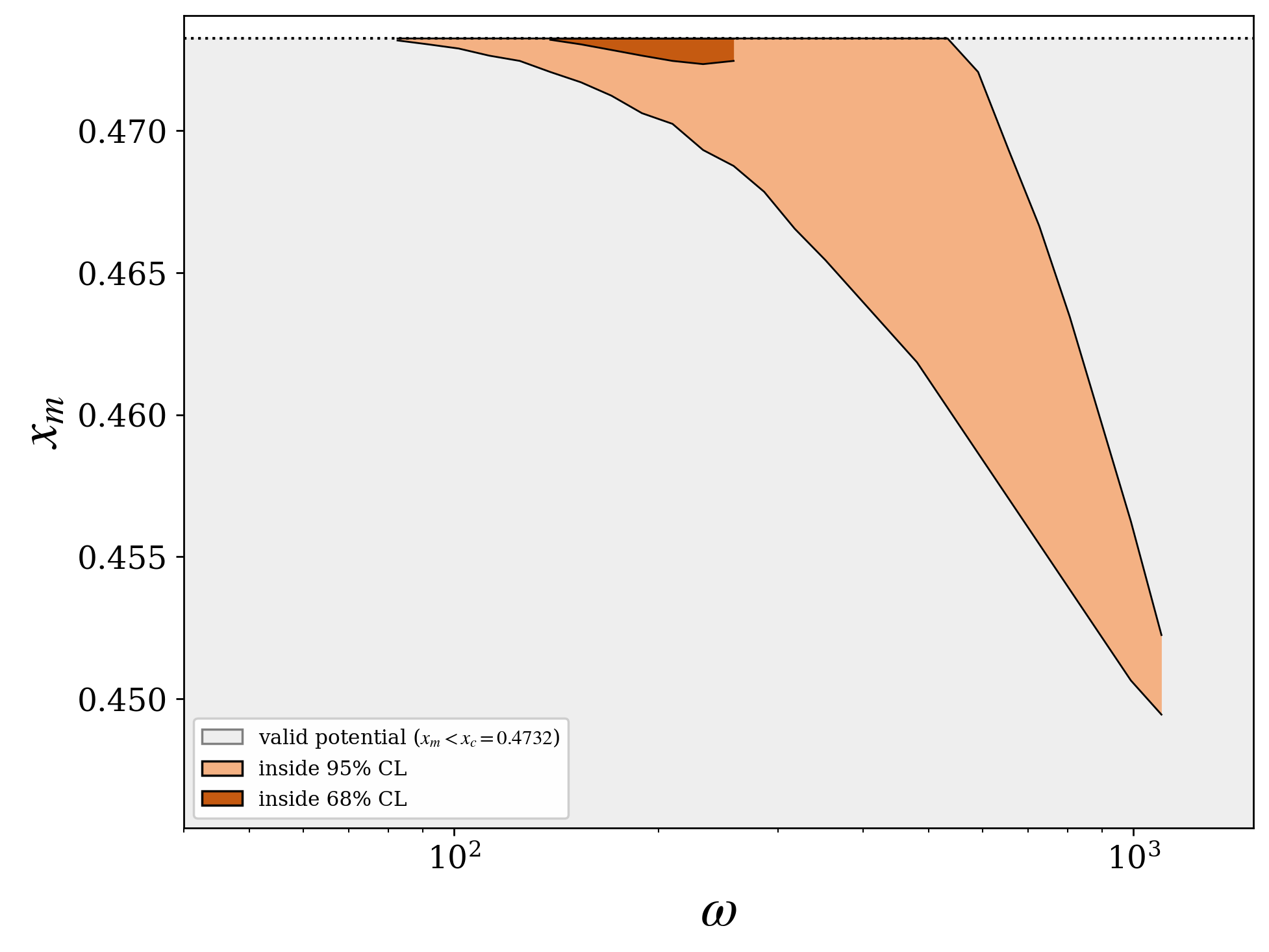}
	\caption{Ranges of the parameters $ \omega $ and $ x_m$ plane for $n=3$ and $N=70$, for which the predicted $(n_s,\ r)$ lies inside the joint 68\% CL (dark) and 95\% (light) CL of the
		P-ACT-LB-BK18 datasets. The dotted line marks the critical value $x_c=0.4732$ above which the model has no inflationary solution. The allowed range is
		$77\lesssim\omega\lesssim1148$ at 95\% CL and $134\lesssim\omega\lesssim272$ at 68\% CL; in both cases $x_m$ must be
		close to $x_c$.}\label{fig:region}
\end{figure}

\begin{figure}[H]
	\centering
	\includegraphics[width=0.8\textwidth]{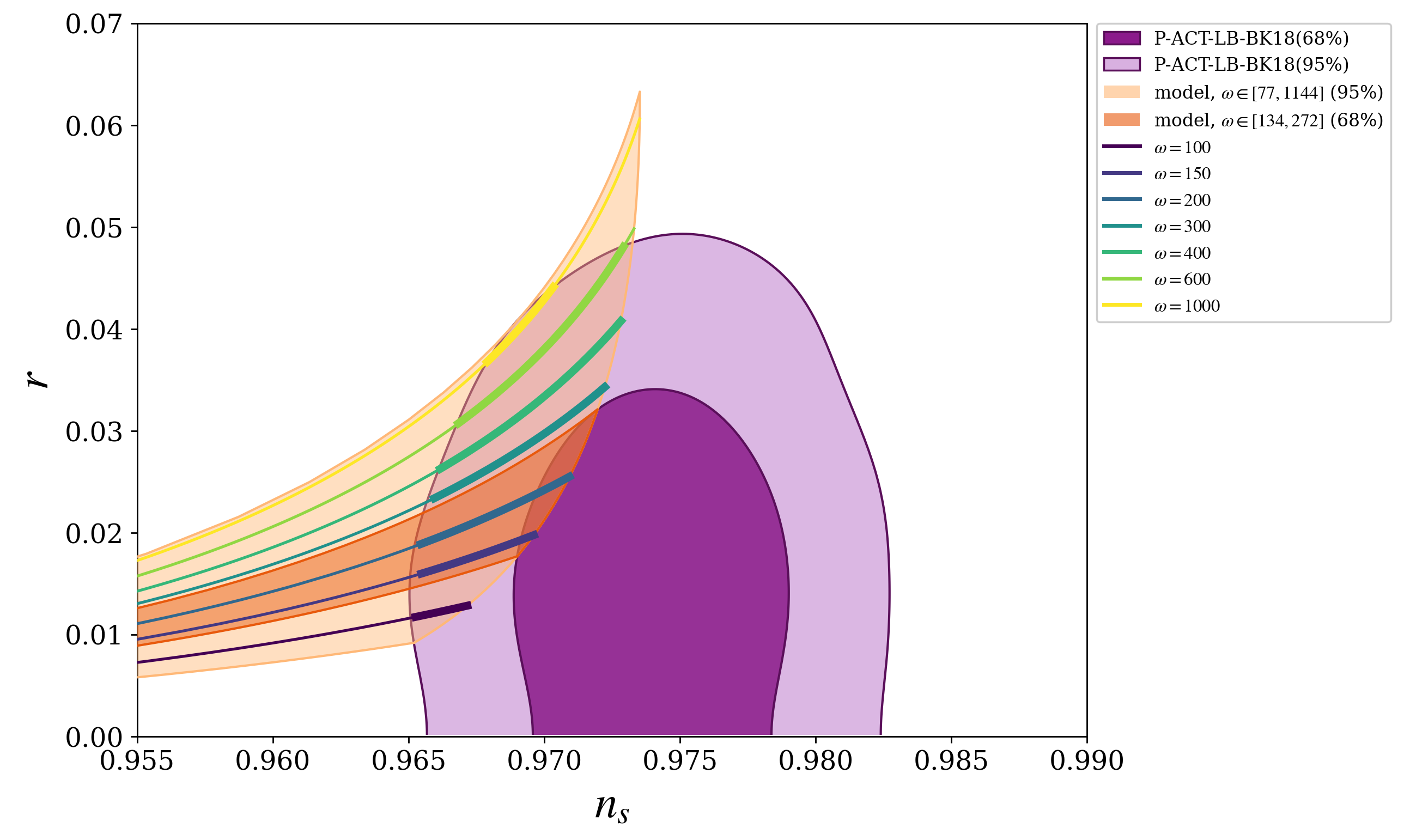}
	\caption{Predictions of the BDU inflationary model in the $n_s$--$r$ plane, compared with the observational constraints from P-ACT-LB-BK18 datasets. Each curve corresponds to a fixed $\omega$ and varying $x_m$; the thick part of a curve lies inside the $ 68\% $ and $ 95\% $ CL. The light orange region shows the predictions of the model for the range of $\omega$ allowed at 95\% CL ($77\lesssim\omega\lesssim1148$) and the dark orange region for the range allowed at $  68\% $ CL ($134\lesssim\omega\lesssim272$).}
	\label{fig.2}
\end{figure}

\begin{figure}[H]
	\centering
	\includegraphics[width=0.8\textwidth]{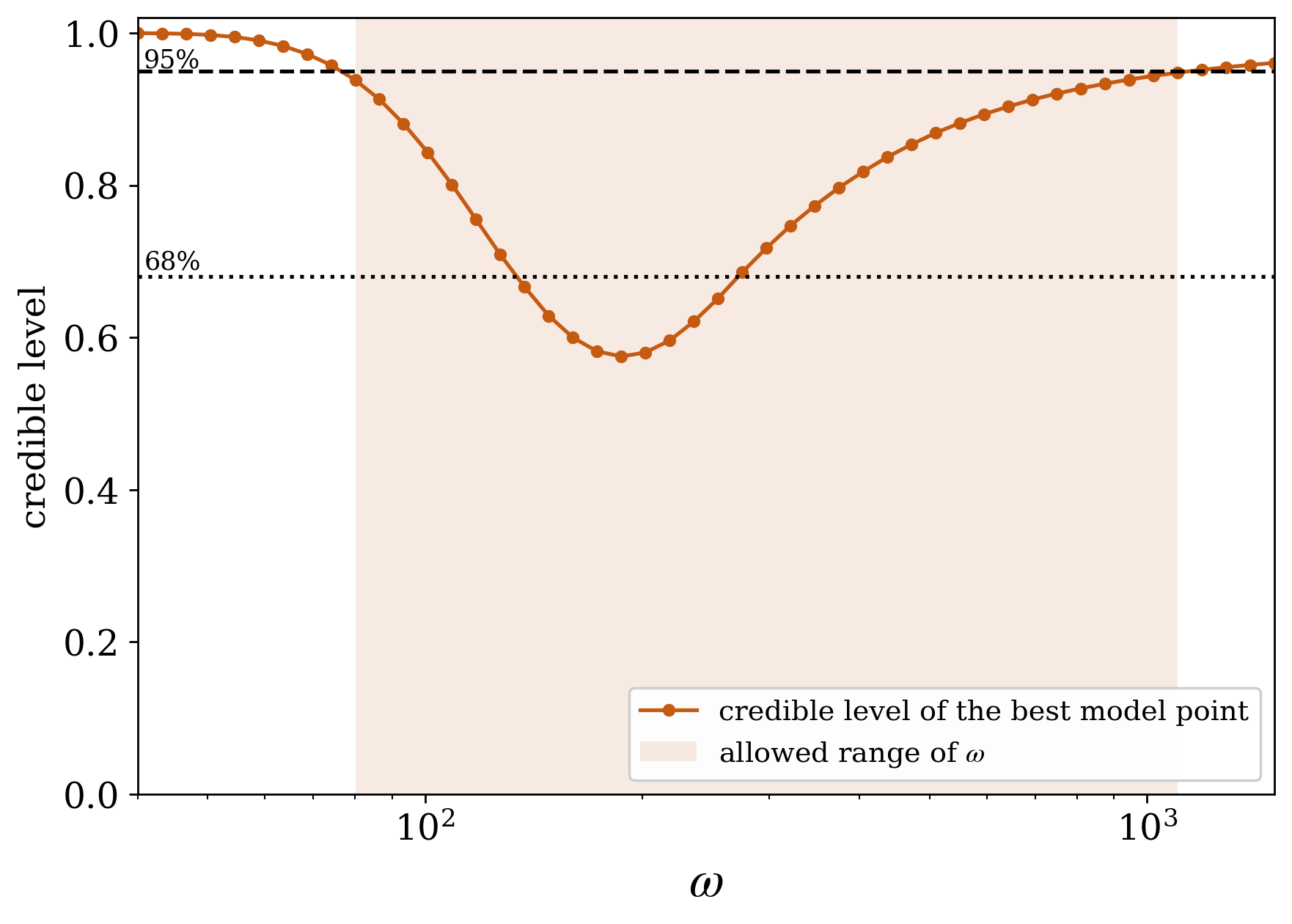}
	\caption{Credible level of the best prediction of the model as a function of $\omega$ ($n=3$, $N=70$). Below the dashed dotted line the prediction lies inside the joint 68\% and 95\% CL. The shaded band is the range of $\omega$ allowed at 95\% CL, $77\lesssim\omega\lesssim1148$; the 68\% CL level is crossed for $134\lesssim\omega\lesssim272$ }
	\label{fig.3}
\end{figure}

In this section, we compare the theoretical predictions of the UBD inflation model with the observational constraints obtained from the P-ACT-LB-BK18 combination in the ACT DR6 analysis with a Gaussian kernel of width $0.55\,\sigma$ \cite{ACT:2025tim, ACT:2025fju}. We analyze the allowed regions of the free parameter space consisting $ n $, $ \omega $, and \(x_m\). To investigate the observational viability of the model, we impose the ACT constraint on the scalar spectral index, \(n_s=0.9741\pm0.0033\), together with the upper bound on the tensor-to-scalar ratio \(r<0.038\) at $ 95\% $ CL. Accordingly, the allowed interval is \(0.9676\leq n_s\leq0.9805\) and \(r<0.038\) at $ 95\% $ CL, and \(0.9689\leq n_s\leq0.9790\) and \(r<0.022\) at $ 68\% $ CL. The aim of this work is to find new and tight constraints on the the BD parameter $\omega$. Therefore, we use the observational values of the $n_s$ and $r$ with the P-ACT-LB-BK18 datasets. For $n=3$ and $N=70$, we find a constraint on the
model’s parameters for $\omega\simeq134$--$272$ and $ x_m \simeq 0.4723$--$ 0.4732 $ at $ 68\%  $CL, and $\omega\simeq77$--$1148$ $ x_m \simeq 0.4494$--$0.4732$ at $ 95\% $ CL. We summarize our numerical study in Table \eqref{tab:ranges} and Table \eqref{tab2}.
Figure~\eqref{fig:region} shows the allowed region in the $(\omega,\ x_m)$ plane, where $x_c$ is the largest value of $x_m$ for which slow--roll inflation is possible (for $x_m>x_c$ the parameter $\epsilon_3$ never vanishes); the allowed region lies close to $x_c$, so $x_m$ must be tuned to it. In Figure~\eqref{fig.2}, the resulting trajectories in the $n_s$--$ r $ plane are illustrated. Each curve corresponds to a fixed value of the BD parameter $\omega$, and moves along the curve as $x_m$ approaches its critical value $x_c$. The light orange region covers all predictions of the model for $\omega$ in the range allowed at $ 95\% $ CL, $77\lesssim\omega\lesssim1148$, and the dark orange region those for $134\lesssim\omega\lesssim272$, the range allowed at $ 68\% $ CL. 
The thick parts of the curves indicates consistency with the data. Figure~\eqref{fig.3}  summarises the comparison with the data as a function of $\omega$. This figure the credible level of the model prediction with the highest posterior density in the $n_s$--$r$ plane. It reaches a minimum of $58\%$ CL at $\omega\simeq190$.  Moreover, it remains below $ 95\%  $CL for $ 77\lesssim\omega\lesssim1148 $ and below $ 68\% $ CL for $ 134\lesssim\omega\lesssim272 $. These intervals determine the observationally allowed range of $ \omega $.

\begin{table}[htbp]
	\centering
	\caption{Parameter ranges for which the predicted $(n_s,\ r)$ lies inside the 68\% and 95\% CL credible regions of the P-ACT-LB-BK18 datasets.}
	\label{tab:ranges}
	\begin{tabular}{lcc}
		\toprule
		Quantity & 68\% CL & 95\% CL \\
		\midrule
		$\omega$ & $134$ -- $272$ & $77$ -- $1148$ \\
		$x_m=\phi_m/\phi_0$ & $0.4723$ -- $0.4732$ & $0.4494$ -- $0.4732$ \\
		$x_{\rm hc}=\phi_{\rm hc}/\phi_0$ & $0.183$ -- $0.219$ & $0.159$ -- $0.299$ \\
		$x_e=\phi_e/\phi_0$ & $0.426$ -- $0.437$ & $0.415$ -- $0.447$ \\
		\bottomrule
	\end{tabular}
\end{table}

\begin{table}[htbp]
	\centering
		\caption{Benchmark parameter sets ($n=3$, $N=70$) for each $\omega$, the point of the family ($x_m$ varied) with the highest joint posterior density of the P-ACT-LB-BK18 datasets. The three points lie inside the 68\% and 95\% CL, respectively.}
	\small
		\label{tab2}
	\begin{tabular}{rcccccccc}
		\toprule
		$\omega$ & $x_m$ & $c$ & $\hat\Lambda$ & $x_{hc}$ & $x_e$ & $\epsilon_1$ & $\epsilon_2$ & $\epsilon_3$  \\
		\midrule
		200  & 0.4732 & 0.4721 & $-0.2500$ & 0.2031 & 0.4327 & $-0.00040$ & 0.0173 & 0.00199 \\
		500  & 0.4670 & 0.4741 & $-0.2502$ & 0.2588 & 0.4403 & $0.00081$  & 0.0146 & 0.00153  \\
		1000 & 0.4530 & 0.4783 & $-0.2506$ & 0.2947 & 0.4343 & $0.00140$  & 0.0137 & 0.00112  \\
		\bottomrule
	\end{tabular}

\end{table}

\subsection{Determination of the Potential Coupling Parameter $\lambda$}
\begin{figure}[htbp]
	\centering
	\includegraphics[width=0.48\textwidth]{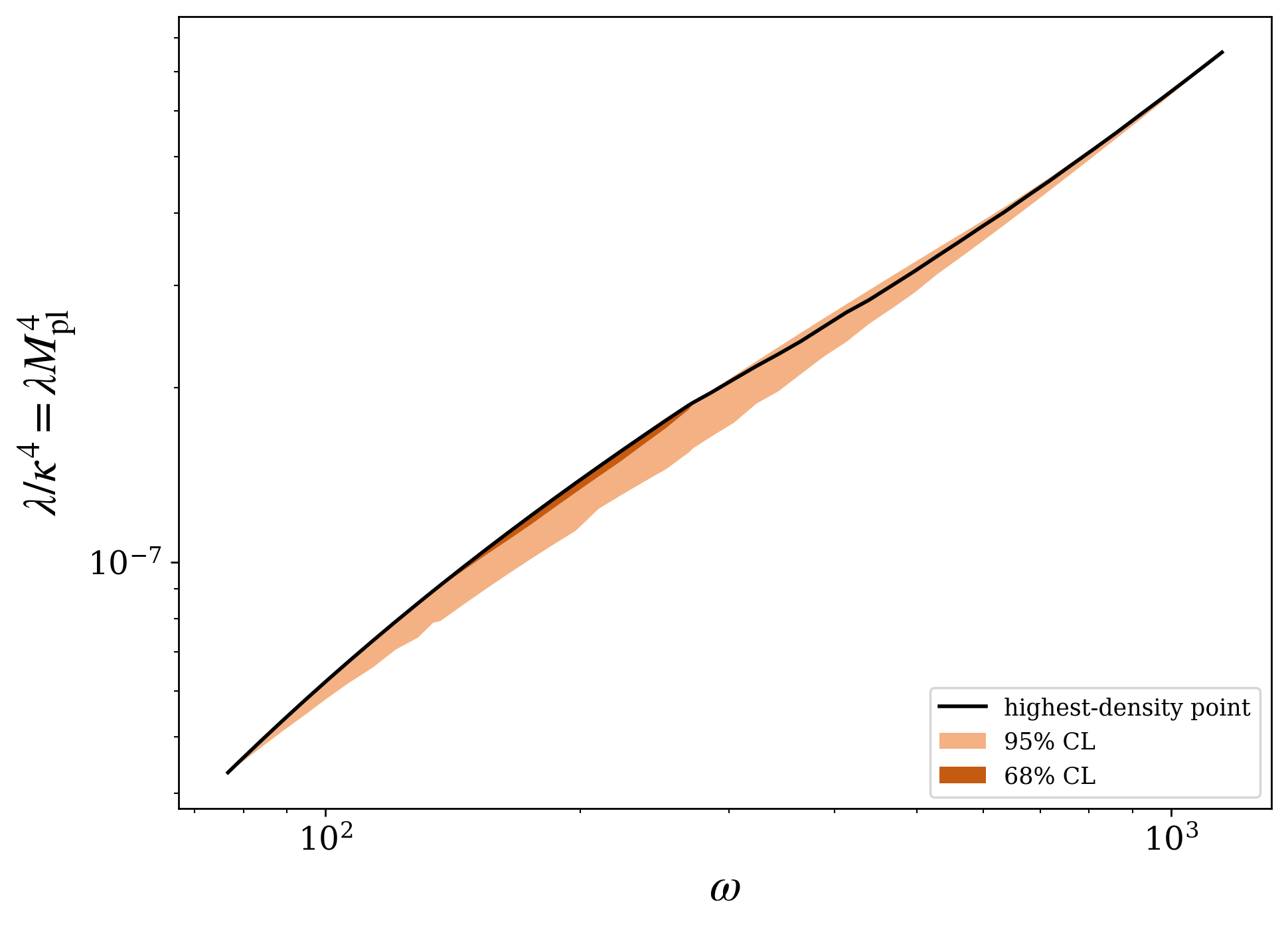}
	\includegraphics[width=0.48\textwidth]{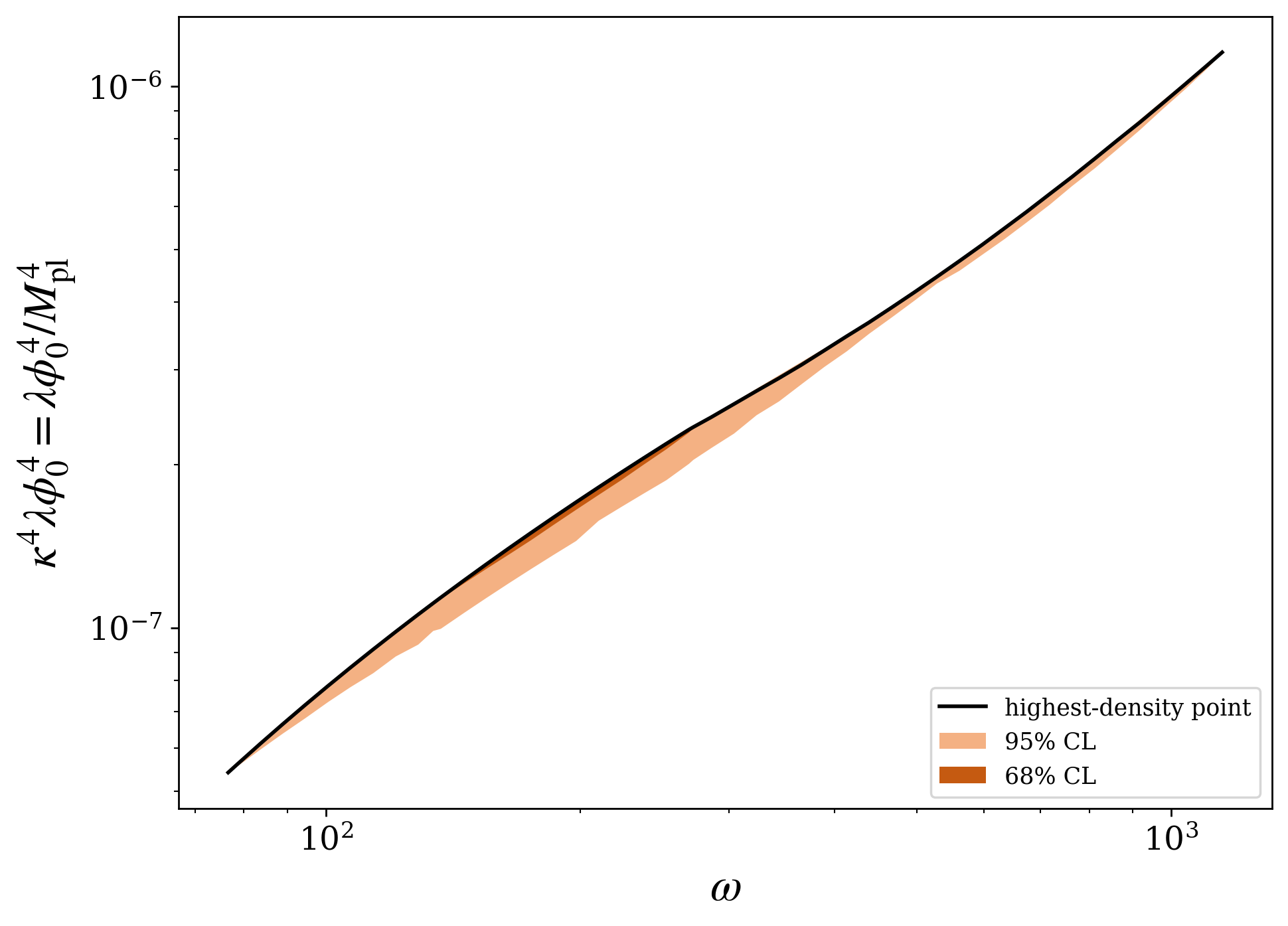}
	\caption{The coupling $\lambda/\kappa^{4}$ (left) and the energy scale $\kappa^{4}\lambda\phi_0^{4}$ (right) as functions of $\omega$ for $n=3$, $N=70$ and $\mathcal{A}_s=2.1\times10^{-9}$. The light and dark bands show the ranges allowed at 95\% and 68\% CL.}
	\label{fig:lambda}
\end{figure}

Having obtained the allowed range of $\omega$ and $x_m$, we now use the amplitude of the primordial spectra to fix the remaining parameter of the potential, the coupling $\lambda$ of the CW term. 
For the action $\int\!\sqrt{-g}\,\phi R$ the effective Planck mass is $M_{\rm eff}^{2}=2\phi$ (it reduces to $M_{\rm pl}^{2}=1/\kappa^{2}$ for $\phi=\phi_{G}=1/2\kappa^{2}$), so the tensor amplitude at horizon crossing is
$\mathcal P_t=2H_{hc}^{2}/(\pi^{2}M_{\rm eff}^{2})=H_{hc}^{2}/(\pi^{2}\phi_{hc})$. With $\mathcal P_t=r\mathcal A_s$, we obtain
\begin{equation}
	H_{hc}^{2}=\pi^{2}\,\phi_{ hc}\,r\,\mathcal A_s \,.
\end{equation}
Requiring that the minimum of the potential reproduces Newton's constant,
$\kappa^{2}\phi_{\rm vac}=1/2$ with $\phi_{\rm vac}=x_m\phi_0$, fixes \ $\phi_0=1/(2\kappa^{2}x_m)$ and gives
\begin{equation}
		H_{hc}= \frac{\pi}{\kappa}\sqrt{\frac{x_{hc}\,r\,A_s}{2x_m}}\,,
\end{equation}
Using Eq.~\eqref{eq41} with $V_{cw}+\Lambda=X\,\phi_{hc} H_{hc}^{2}=\lambda\phi_0^{4}W(x)$, we obtain
 \begin{equation}
 	\frac{\lambda}{\kappa^{4}}=\frac{4\pi^{2}\mathcal A_s\,r\,X\,(x_{\rm hc}x_m)^{2}}{W(x_{\rm hc})}\,, \qquad
 	\kappa^{4}\lambda\phi_0^{4}=\frac{1}{16\,x_m^{4}}\,\frac{\lambda}{\kappa^{4}}\,.
 \end{equation}
Because $\phi$ has dimension (mass)$^{2}$ and $V$ has dimension (mass)$^{4}$, $\lambda$ has dimension mass$^{-4}$;
hence $\lambda/\kappa^{4}=\lambda M_{\rm pl}^{4}$ is dimensionless ($M_{\rm pl}$ is the reduced Planck mass,
$\kappa^{2}=M_{\rm pl}^{-2}$), and $\lambda\phi_0^{4}$ in units of $M_{\rm pl}^{4}$ gives the energy scale of the potential.
We use $\epsilon_3\sim10^{-3}$ for the allowed parameters, $X\simeq6.01$--$6.02$, and  $\mathcal A_s=2.1\times10^{-9}$, $N=70$ and $n=3$. 
Over the whole region allowed by the data the coupling lies in
\begin{equation}
	\frac{\lambda}{\kappa^{4}}\simeq(0.89\text{--}1.9)\times10^{-7}\ \ (68\%),\qquad
	\frac{\lambda}{\kappa^{4}}\simeq(0.43\text{--}76)\times10^{-7}\ \ (95\%),
\end{equation}
corresponding to $\kappa^{4}\lambda\phi_0^{4}\simeq(1.1\text{--}2.4)\times10^{-7}$ and
$(0.54\text{--}12)\times10^{-7}$, respectively. The value of $\lambda$ increases with $\omega$ because $r$ and
$x_{\rm hc}$ grow with $\omega$ along the allowed region. Figure~\eqref{fig:lambda} shows $\lambda/\kappa^{4}$ (left) and $\kappa^{4}\lambda\phi_0^{4}$ (right) as functions of $\omega$ over the allowed range: both grow almost linearly with $\omega$.

The value of $\lambda$ scales approximately linearly with $\omega$ and depends on the normalization $\kappa^{2}\phi_{\rm vac}=1/2$ (the physical vacuum value of $\phi$); with $\kappa^{2}\phi_{0}=1/2$ instead, it would be larger by $1/x_m^{2}\simeq4$--$5$.\\

%#############################################################

\section{Discusion and Conclusions}\label{sec5}

In this paper, we have studied the dynamics of BD gravity theory within the framework of unimodular gravity with the CW potential. This combination inspires a new approach
to both the inflationary phase of the early universe and providing a solution to the cosmological constant problem and so is a good proposal to resolve the cosmological constant problem (fine--tuning). Unimodular gravity can be viewed as a restricted formulation of GR. In this framework, the cosmological constant is usually treated as an integration constant or, equivalently, as a Lagrange multiplier. By deriving the field equations for this theory under the spatially flat FRW metric with the unimodular constraint, we find analytic solution of the field at slow--roll inflation regime. For the purposes of our analysis, we consider the Lagrange multiplier as a power law ansatz $\Lambda(\phi)=\Lambda_0+A\,\phi^{n}$. Data analysis in the parameter space shows some permissible valid values for the theoretical parameters of the solutions by according to the observational data, where the latest observational constraints are used from ACT collaborations joint data. In this study, we obtained observationally consistent ranges on the parameter $ \omega $. We also find that the inflationary observables predicted by our model are consistent with the observational constraints within the slow--roll regime. This agreement with observational data supports the viability of the proposed inflationary model. The comparison with the P-ACT-LB-BK18 joint data gives the constraint $77\lesssim\omega\lesssim1148$ at 95\% CL and $134\lesssim\omega\lesssim272$ at 68\% CL for $n=3$ and $N=70$. Therefore, the UBD model with the CW potential provides a viable and observationally consistent cosmological inflationary model. 

We have fixed the number of e-folds to $N=70$. Relating $N$ to the post-inflationary history requires a model of reheating. So $N=70$ is somewhat larger than the typical range expected, and the results should be read as conditional on a non-standard post-inflationary expansion, whose study we defer to a forthcoming work. For future investigation, we propose a detailed analysis of the reheating phase following the end of inflation in this model.

\section{Declaration of competing interest}
The authors declare that they have no known competing financial
interests or personal relationships that could have appeared to
influence the work reported in this paper.
\section{Funding Declaration}
This research is supported by the postdoc. grant of the Semnan
University (number 20251143).
\section{Data availability}
Data which we used in this work are addressed in reference section.

\end{document}